\documentclass[twocolumn,showpacs,preprintnumbers,amsmath,amssymb,nofootinbib]{revtex4}
\usepackage{graphicx}
\usepackage{dcolumn}
\usepackage{bm}
\usepackage{amsmath}    
\usepackage{graphicx}   
\newcommand{\be}{\begin{equation}}
\newcommand{\ee}{\end{equation}}
\newcommand{\bea}{\begin{eqnarray}}
\newcommand{\eea}{\end{eqnarray}}

\begin{document}

\title{Nuclear response theory with multiphonon coupling in a covariant framework}

\author{Elena Litvinova}
\affiliation{Department of Physics, Western Michigan University,
Kalamazoo, MI 49008-5252, USA } \affiliation{National
Superconducting Cyclotron Laboratory, Michigan State University,
East Lansing, MI 48824-1321, USA}
\date{\today}
\begin{abstract}
\ \ \
\\
{\bf Background:} Nuclear excited states within a wide range of
excitation energies are formally described by the linear response
theory. Besides its conventional formulation within the
quasiparticle random phase approximation (QRPA) representing excited
states as two correlated quasiparticles (2q), there exist extensions
for 4q configurations. Such extended approaches are quite successful
in the description of gross properties of nuclear spectra, however,
accounting for many of their fine features requires further
extension
of the configuration space.\\
{\bf Purpose:}  This work aims at the development of an approach which
is capable of such an extension as well as of reproducing and predicting
fine spectral properties, which are of special interest at low energies. \\
{\bf Method:} The method is based on the covariant density
functional theory and time blocking approximation, which is
extended for couplings between quasiparticles and multiphonon excitations. \\
{\bf Results:} The covariant multiphonon response theory is
developed and adopted for nuclear structure calculations in
medium-mass and heavy nuclei. The equations are formulated in both general
and coupled forms in the spherical basis. \\
{\bf Conclusions:} The developed covariant multiphonon response
theory represents a new generation of the approaches to nuclear
response, which aims at a unified description of both high-frequency
collective states and low-energy spectroscopy
in medium-mass and heavy nuclei.
\end{abstract}


\maketitle

\section{Introduction}

Linear response theory is a conventional framework adopted for
calculations of nuclear spectra in the low-energy regime, i.e. at
excitation energies below and around $\sim$ 100 MeV. Such
excitations represent the response of nuclear systems to
sufficiently weak external fields, so that the induced changes of
the nucleonic density are small compared to the ground state density
and can be treated in the linear approximation. The self-frequencies
of the oscillations of the nucleonic densities can be found as
solutions of the secular equation in the limit of the vanishing
external field. The simplest non-perturbative approach to the
secular equation for strongly interacting Fermions includes
scattering process of particle-hole pairs which can be described by
ring diagrams summed up to the infinite order. This approach is
known as the random phase approximation (RPA) \cite{BP.53} or the
quasiparticle random phase approximation (QRPA)
\cite{Bo.59,Ba.60,So.71}, where the latter is a generalization of
the former to the superfluid case and since 1960's has become a
standard approach to vibrational spectra of nuclei. Besides its
typical diagrammatic structure, the QRPA calculations are based on
the concept of nucleon-nucleon (effective) residual interaction
which has evolved considerably over the years. The use of relatively
simple multipole-multipole forces \cite{So.71} and Landau-Migdal
interaction \cite{Mig.67} allowed for reasonable explanation of some
experimental findings although the agreement with data could only be
achieved after fine tuning of the interaction parameters.

Over the decades, various approaches to the nucleon-nucleon residual
interaction, such as G-matrix \cite{B.55,G.57,BBP.63}, Skyrme
\cite{S.59,VB.72}, Gogny \cite{DG.80} or Fayans \cite{STF.88}
interactions have been developed and successfully tested on nuclear
structure calculations. The relativistic approach, based on the
Walecka model \cite{Wal.74,SW.86,Rin.96} for meson-exchange
nucleon-nucleon interaction, has become very successful after the
inclusion of non-linear meson coupling \cite{BB.77} or the density
dependence of the coupling vertices
\cite{TW,DDME1,DDME2,PRN.03,RVCRS.11}, see also review
\cite{VALR.05} and references therein.

The progress in computer technologies has allowed for fast execution
of complex numerical algorithms and for self-consistent QRPA
calculations with the above mentioned interactions, in contrast to
the earlier ones with simple effective interactions which were
disconnected from the underlying mean-field. At the same
time, approaches beyond QRPA were developed to account for effects
of more complex nature than particle-hole (1p1h) or
two-quasiparticle (2q) configurations to overcome the principal
limitation of the QRPA in the description of the nuclear response.

Medium-mass and heavy nuclei represent Fermi-systems where
single-particle and vibrational degrees of freedom are strongly
coupled. Collective vibrations lead to shape oscillations of the
mean nuclear potential and, therefore, modify the single-particle
motion. To take this effect into account, already in Ref.
\cite{BM.75} a general concept for the quasiparticle-vibration
(phonon) coupling (QVC) part of the single-nucleon self-energy has
been proposed. This concept had various implementations over the
years within the Quasiparticle-Phonon Model (QPM)
\cite{SSV.82,GSV.88,Sol.92}, Nuclear Field Theory
\cite{BBBL.77,BBB.83,BBBD.85,BB.81,CBGBB.92,CB.01,SBC.04} and others
\cite{BG.80,AK.99,BBG.99,B.09,CSFB.10,Tse.89,DNSW.90,KTT.97,KST.04,Tse.07,LT.07}.
In particular, the approaches \cite{Tse.89,KTT.97,Tse.07,LT.07} are
formulated as a theory for nuclear response function. These studies,
however, are either not self-consistent or do not include pairing
correlations of the superfluid type. Recently, a set of
self-consistent approaches to QVC in the relativistic framework has
become available
\cite{LR.06,LA.11,LRT.07,LRT.08,LRT.10,L.12,LRT.13}, where the
latter four account for the superfluid pairing. It has been shown
that these models improve considerably the description of the
single-particle states around the Fermi surface and explain the
strong fragmentation of deep hole states, giant resonances and soft
modes quantitatively with a good precision, despite the very limited
number of parameters in the underlying Lagrangian.

The present work focuses on the nuclear response theory which
includes QVC and superfluid pairing on equal footing. Specifically,
the nuclear response in the particle-hole channel, which describes a
large variety of typical nuclear excited states, is considered.
Based on the relativistic quasiparticle time blocking approximation
(RQTBA) developed in Ref. \cite{LRT.08} and its two-phonon version
\cite{LRT.10,LRT.13}, it is shown how higher-order QVC effects, or
multiphonon coupling, can be included self-consistently. The
approach is formulated as a non-perturbative extension of the RQTBA.
The convergence of the response function with respect to the number
of coupled phonon modes is justified in terms of its multipole
expansion in the spherical basis.

It is implied that the response theory with multiphonon coupling
presented here is based on the relativistic description of the
nuclear uncorrelated ground state known as the covariant density
functional theory (CDFT), although the approach can be
straightforwardly adopted for calculations with other types of
underlying density functionals, such as Skyrme, Gogny, Fayans etc.

\section{Relativistic quasiparticle time blocking approximation: a brief overview}

This section introduces nuclear response formalism and reviews the
relativistic time blocking approximation, which serves as a
foundation for the extended approach. To maintain consistency with
the previous versions of RQTBA, the notations are kept close to
those of Ref. \cite{LRT.13}.

The response function of a finite Fermi-system with an even particle
number describes propagation of two quasiparticles in the medium and
quantifies the response of the system to an external perturbation.
The exact propagator includes, ideally, all possible kinds of the
in-medium interaction between two arbitrary quasiparticles and
contains all the information about the Fermi system, which can be,
in principle, extracted by a certain experimental probe, if its
interaction with the system is represented by a single-quasiparticle
operator.

In the case of a weak external field, the response function $R$ is
conventionally described by the Bethe-Salpeter equation (BSE). The
general form of this equation
\bea
R(14,23) = G(1,3)G(4,2) - \nonumber\\
- i\sum\limits_{5678}G(1,5)G(6,2)U(58,67)R(74,83),
\label{bse0}%
\eea
includes the one-nucleon Green function (propagator) $G(1,2)$ in the
nuclear medium and the effective nucleon-nucleon interaction
$U(14,23)$ irreducible in the relevant channel. Here and below the
particle-hole channel is considered. For the systems with pairing
correlations of the superfluid type the conventional degrees of
freedom are quasipaticles in Bogoliubov's sense represented by
superpositions of particles and holes on top of the Hartree (or
Hartree-Fock) Fermi sea. To account for the superfluidity effects,
we use the formalism of the extended (doubled) space of
quasiparticle states described in Refs.~\cite{Tse.07,LRT.08}. Thus,
the generic number indices $1,2,...$ include all
single-quasiparticle variables in an arbitrary representation,
components in this doubled space, and time. Respectively, the
summation over the number indices implies an integration over the
time variables. The amplitude $U$ is determined as a variational
derivative of the nucleonic self-energy $\Sigma$ with respect to the
exact single-quasiparticle Green function $G$:
\begin{equation}
U(14,23)=i\frac{\delta\Sigma(4,3)}{\delta G(2,1)}. \label{uampl}%
\end{equation}
If the ground state can be with a reasonable accuracy described by a
static mean-field, it is convenient to decompose both the
single-quasiparticle self-energy $\Sigma$ and the irreducible
effective interaction $U$ into static ${\tilde\Sigma}, \tilde V$ and
time-dependent (energy-dependent) $\Sigma^{(e)}, U^{(e)}$ parts as
\bea
\Sigma = {\tilde\Sigma} + \Sigma^{(e)}\\
U = {\tilde V} + U^{(e)}.
\eea Accordingly, the uncorrelated response is introduced as
${\tilde R}^{(0)}(14,23)={\tilde G}(1,3){\tilde G}(4,2)$, where
${\tilde G}(1,2)$ are the single-quasiparticle mean-field Green
functions in the absence of the term $\Sigma^{(e)}$ in the
self-energy. The Green functions $G$ and $\tilde G$ are connected by
the Dyson equation:
\be G(1,2) = {\tilde G}(1,2) + \sum\limits_{34}{\tilde
G}(1,3)\Sigma^{(e)}(3,4)G(4,2), \label{dyson} \ee
%
so that $G$ can be eliminated from Eq. (\ref{bse0}) and, after some
simple algebra, the BSE (\ref{bse0}) takes the form:
\bea
R(14,23) = {\tilde G}(1,3){\tilde G}(4,2) - \nonumber\\
- i\sum\limits_{5678}{\tilde G}(1,5){\tilde G}(6,2)V(58,67)R(74,83),
\label{bse1}%
\eea
where $V$ is the new effective interaction amplitude which is
specified below. The well-known quasiparticle random phase
approximation QRPA including its relativistic version (RQRPA)
corresponds to the case of $V={\tilde V}$ neglecting the
time-dependent term $U^{(e)}$. More precisely, in the (R)QRPA the
time-dependent term is included in a static approximation by
adjusting the parameters of the effective interaction $\tilde V$ to
ground state properties of nuclei such as masses and radii. In the
self-consistent (R)QRPA the static effective interaction is the
second variational derivative of the covariant energy density
functional (CEDF) $E[{\cal R}]$ with respect to the density matrix
$\cal R$ \cite{VALR.05}:
\be
{\tilde V}(14,23) = \frac{2\delta^2 E[{\cal R}]}{\delta{\cal
R}(2,1)\delta{\cal R}(3,4)} . \label{2-deriv}
\ee

In the approaches beyond the QRPA both static and time-dependent
terms are contained in the residual interactions. In medium-mass and
heavy nuclei vibrational and rotational modes are strongly coupled
to the single-particle ones. In particular, the coupling to
low-lying vibrations is known already for decades \cite{BM.75} as a
very important mechanism of the formation of nuclear excited states
and serves as a foundation for the so-called (quasi)particle-phonon
coupling model. Implementations of this concept on the base of the
modern density functionals have been extensively elaborated in
non-relativistic
\cite{CB.01,TSG.07,TTKG.07,TSK.09,CSB.10,NCBBM.12,NCV.14} and
relativistic
\cite{LR.06,LA.11,LRT.07,LRT.08,LRT.10,L.12,LRT.13,MLVR.12,LBFMZ.14}
frameworks.

\begin{figure}
\begin{center}
\includegraphics[scale=0.4]{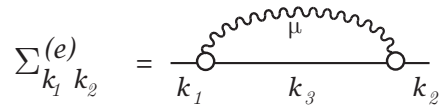}
\end{center}
\caption{Top: The skeleton graph representing the
single-quasiparticle self-energy $\Sigma^{(e)}$. The solid straight
line denotes the single-quasiparticle nucleonic propagator, the
Latin indices stand for the single-quasiparticle quantum numbers,
the wiggly line with Greek index shows the phonon propagator with
phonon quantum numbers, empty circles represent QVC vertices.}
\label{fig0}%
\end{figure}

Going beyond the Hartree (Hartree-Fock) approach of the CDFT, it is
natural to include non-perturbatively bubble and ladder types of
nucleon-nucleon correlations associated with multiple meson exchange
and re-scattering. This becomes possible because these processes
lead to the emergence of collective effects of vibrational
character. These vibrations (phonons) manifest themselves as the new
degrees of freedom associated with the new order parameter
corresponding to the quasiparticle-vibration coupling vertices,
which helps to classify and decouple different correlations in the
self-consistent non-perturbative treatment. For instance, one-phonon
exchange is the leading-order approximation for the time-dependent
parts of the effective interaction $U^{(e)}$ and of the nucleonic
self-energy $\Sigma^{(e)}$, whose Fourier transform to the energy
domain
\be \Sigma^{(e)}(1,2;\varepsilon) =
\sum\limits_{34}\int\limits_{-\infty}^{+\infty}\frac{d\omega}{2\pi
i}\Gamma^{(e)}(14,23;\omega) G(3,4;\varepsilon+\omega)
\label{sigmae} \ee is formally expressed as a convolution of the
exact single-quasiparticle Green function $G$ and two-quasiparticle
scattering amplitude $\Gamma^{(e)}$. In the leading-order
approximation with respect to the QVC, $\Gamma^{(e)}$ is
obtained as an infinite sum of the ring diagrams with meson-exchange
interaction. Thus, $\Sigma^{(e)}$ is separated from Hartree-Fock
contributions which are supposed to be included in the static
self-energy $\tilde\Sigma$. The dynamical part of the self-energy
can be represented by the Feynman graph shown in Fig. \ref{fig0},
where in the leading order the straight line stands for the
mean-field single-nucleon propagator $G={\tilde G}$ and the wiggly
line replaces an infinite sum of the ring diagrams.

To account for higher-order correlations, after solution of the
Dyson equation (\ref{dyson}) the obtained Green function, together
with the amplitude $\Gamma^{(e)}$ calculated from the Eq.
(\ref{bse1}), for instance, in the approach described below, can be
substituted to Eq. (\ref{sigmae}), thus dressing the skeleton graph
shown in Fig. \ref{fig0}.

The consistent response formalism for the BSE (\ref{bse1}) based on
the QVC self-energy of Eq. (\ref{sigmae}) has become possible in the
time blocking approximation, first proposed in Refs.
\cite{Tse.07,LT.07} for superfluid Fermi systems and elaborated in
Ref. \cite{LRT.08} in detail for the relativistic framework. This
approximation allows for exact summation of a selected class of
Feynman's diagrams which give the leading contribution of the
quasiparticle-phonon coupling effects to the response function.
Following \cite{LRT.08}, it is convenient to write the BSE
(\ref{bse1}) in the representation in which the mean-field Green
function $\tilde{G}$ is diagonal. This representation is given by
the set of the eigenfunctions $|\psi_{k}^{(\eta)}\rangle$ of the
Relativistic Hartree-Bogoliubov (RHB) Hamiltonian
$\mathcal{H}_{RHB}$ satisfying the equations \cite{KuR.91}:
\begin{equation}
\mathcal{H}_{RHB}|\psi_{k}^{(\eta)}\rangle=\eta E_{k}|\psi_{k}^{(\eta)}%
\rangle,\ \ \ \ {\cal H}_{RHB} = 2 \frac{\delta E[{\cal
R}]}{\delta{\cal R}},
\label{hb}%
\end{equation}
where $E_{k}>0$, the index $k$ stands for the set of the
single-particle quantum numbers including states in the Dirac sea,
and the index $\eta = \pm 1$ labels positive- and negative-frequency
solutions of Eq.~(\ref{hb}) in the doubled quasiparticle space. The
eigenfunctions $|\psi_{k}^{(\eta)}\rangle$ are 8-dimensional
Bogoliubov-Dirac spinors:
\begin{equation}
|\psi_{k}^{(+)}(\mbox{\boldmath $r$})\rangle=\left(
\begin{array}
[c]{c}%
U_{k}(\mbox{\boldmath $r$})\\
V_{k}(\mbox{\boldmath $r$})
\end{array}
\right)  ,\ \ \ \ |\psi_{k}^{(-)}(\mbox{\boldmath
$r$})\rangle=\left(
\begin{array}
[c]{c}%
V_{k}^{\ast}(\mbox{\boldmath $r$})\\
U_{k}^{\ast}(\mbox{\boldmath $r$})
\end{array}
\right)  , \label{dbasis}%
\end{equation}
which form the working basis called Dirac-Hartree-BCS (DHBCS) basis
for the subsequent calculations.

Within the time blocking approximation and after performing a
Fourier transformation to the energy domain, the BSE (\ref{bse1})
for the spectral representation of the nuclear response function
$R(\omega)$ in the basis $\{|\psi_{k}^{(\eta)}\rangle\}$ reads:
\bea
R_{k_{1}k_{4},k_{2}k_{3}}^{\eta\eta^{\prime}}(\omega)=\tilde{R}_{k_{1}k_{2}%
}^{(0)\eta}(\omega)\delta_{k_{1}k_{3}}\delta_{k_{2}k_{4}}\delta^{\eta
\eta^{\prime}}+\nonumber\\
+ \tilde{R}_{k_{1}k_{2}}^{(0)\eta}(\omega)
\sum\limits_{k_{5}k_{6}\eta^{\prime\prime}}{V}_{k_{1}k_{6}%
,k_{2}k_{5}}^{\eta\eta^{\prime\prime}}(\omega)R_{k_{5}k_{4},k_{6}k_{3}}%
^{\eta^{\prime\prime}\eta^{\prime}}(\omega),
\label{respdir}%
\eea
being a matrix equation in the DHBCS basis for each external energy
variable $\omega$. This is the main result of the time-blocking
approximation which separates the integrations over the intermediate
energy variable in such a way that it is fully integrated out in the
interaction amplitude $V$. The quantity ${\tilde R}^{(0)}$
\be
{\tilde R}^{(0)\eta}_{k_1k_2}(\omega) = \frac{1}{\eta\omega -
E_{k_1} - E_{k_2}} \label{mfresp}
\ee
describes the free propagation of two quasiparticles with their
Bogoliubov energies $E_{k_1}$ and $E_{k_2}$ in the relativistic mean
field.
The interaction amplitude of Eq. (\ref{respdir}) contains both
static $\tilde V$ and dynamical (frequency-dependent) $W(\omega)$
parts as follows:
\bea
{V}_{k_{1}k_{4},k_{2}k_{3}}^{\eta\eta^{\prime}}(\omega)=\tilde{V}%
_{k_{1}k_{4},k_{2}k_{3}}^{\eta\eta^{\prime}} + {W}
_{k_{1}k_{4},k_{2}k_{3}}^{\eta\eta^{\prime}}(\omega), \nonumber \\
{W}_{k_{1}k_{4},k_{2}k_{3}}^{\eta\eta^{\prime}}(\omega) = \Bigl[\Phi_{k_{1}k_{4},k_{2}%
k_{3}}^{\eta}(\omega)-\Phi_{k_{1}k_{4},k_{2}k_{3}}^{\eta}(0)\Bigr]\delta%
^{\eta\eta^{\prime}}.
\label{W-omega}%
\eea
\begin{figure*}
\begin{center}
\includegraphics*[scale=0.8]{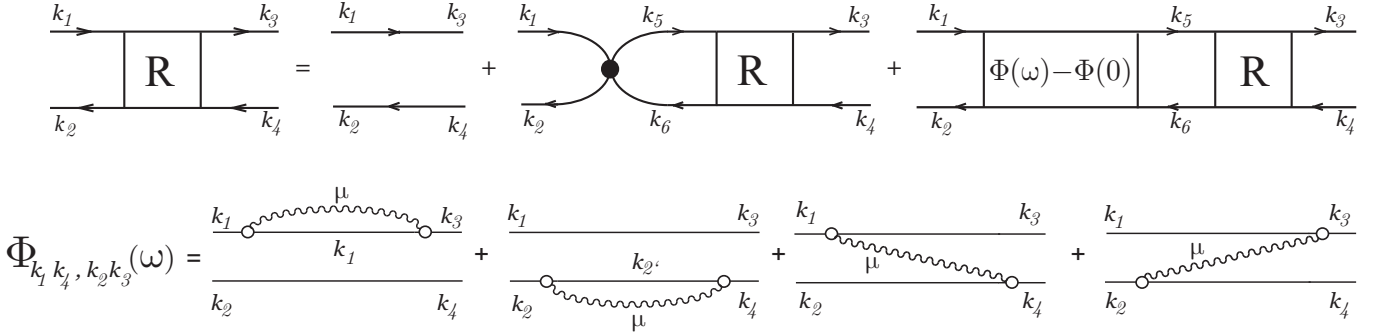}
\end{center}
\caption{Top: Bethe-Salpeter equation for the response function $R$
in the ph-channel in diagrammatic representation. The solid lines
denote single-quasiparticle mean-field propagators. The integral
part is divided into two terms; the small black circle represents
the static effective interaction $\tilde V$  and the
energy-dependent block $\Phi(\omega)-\Phi(0)$ contains the dynamic
contributions. Bottom: The dynamical part of the effective
interaction in the quasiparticle-vibration coupling (QVC) model in
the leading order on QVC. }
\label{fig1}%
\end{figure*}
The diagrammatic representation of the Eq. (\ref{respdir}) with the
interaction of Eq. (\ref{W-omega}) is given in Fig. \ref{fig1}. The
black circle in the second term on the right hand side of the top
line corresponds to the static effective interaction denoted by
${\tilde V}$ and, in the absence of the third term containing the
phonon coupling effects via the amplitude $W$, one would have the
QRPA equation. The energy-dependent resonant part of the
two-quasiparticle amplitude $\Phi (\omega)$ can be factorized
\cite{Tse.07} and takes the following form:
\be
\Phi^{\eta}_{k_1k_4,k_2k_3} (\omega) = %
\sum_{k_5k_6,\mu} \zeta^{\,\mu\eta}_{k_1k_2;k_5k_6}\,
\tilde{R}^{(0)\eta}_{k_5k_6} (\omega - \eta\,\Omega_{\mu})\,
\zeta^{\,\mu\eta *}_{\,k_3k_4;k_5k_6}\,, \label{phires} \ee
so that $\tilde{R}^{(0)\eta}_{k_5k_6}(\omega-\eta\,\Omega_{\mu})$
are the matrix elements of the two-quasiparticle propagator in the
mean field with the frequency shifted forward or backward by the
phonon energy $\Omega_{\mu}$. The quantities $\zeta$ are the
generalized phonon vertices:
\bea
\zeta^{\,\mu(+)}_{\,k_1k_2;k_5k_6} =
\delta^{\vphantom{(+)}}_{k_1k_5}\,\gamma^{(-)}_{\mu;k_6k_2}
-\gamma^{(+)}_{\mu;k_1k_5}\delta^{\vphantom{(+)}}_{k_6k_2},
%
\nonumber \\
\zeta^{\,\mu(-)}_{\,k_1k_2;k_5k_6} =
\delta^{\vphantom{(+)}}_{k_5k_1}\,\gamma^{(+)\ast}_{\mu;k_2k_6}
-\gamma^{(-)\ast}_{\mu;k_5k_1}\delta^{\vphantom{(+)}}_{k_2k_6},
\label{zetapm}
\eea
revealing the four terms in Eq. (\ref{phires}) which correspond to
those in the diagrammatic representation of the amplitude
$\Phi(\omega)$ in the bottom line of Fig. \ref{fig1}.
%
%
%

The shorthand notation for the phonon emission (absorption)
amplitudes imply:
\begin{equation}
\gamma^{\eta}_{\mu;k_1k_2} =
\gamma^{\eta_1\eta_2}_{\mu;k_1k_2}\delta_{\eta\eta_1}\delta_{\eta\eta_2},
\ \ \ \ \eta = (\pm), \label{gammas}
\end{equation}
where ${\gamma}_{\mu;k_{1}k_{2}}^{\eta_{1}\eta_{2}}$ are the matrix
elements of these amplitudes in the doubled quasiparticle space.
They determine the probability of the coupling of a quasiparticle
pair in the states $\{k_1\eta_1\},\{k_2\eta_2\}$ to the collective
vibrational state (phonon) with quantum numbers $\mu =
\{\Omega_{\mu}, J_{\mu}, M_{\mu},\pi_{\mu}\}$.
In the RQTBA these vertices are derived from the corresponding RQRPA
transition densities $\mathcal{R}_{\mu}$ and the static effective
interaction as
\begin{equation}
\gamma_{\mu;k_{1}k_{2}}^{\eta_{1}\eta_{2}}=\sum\limits_{k_{3}k_{4}}%
\sum\limits_{\eta}
\tilde{V}_{k_{1}k_{4},k_{2}k_{3}}^{\eta_{1},%
-\eta_,\eta_{2},\eta}\mathcal{R}_{\mu;k_{3}k_{4}}^{\eta},
\label{phonon}%
\end{equation}
where
$\tilde{V}_{k_{1}k_{4},k_{2}k_{3}}^{\eta_1\eta_4,\eta_2\eta_3}$ is
the matrix element of the amplitude $\tilde{V}$ of
Eq.~(\ref{2-deriv}) in the basis $\{|\psi_{k}^{(\eta)}\rangle\}$.
The matrix elements of the phonon transition densities are
calculated, in first approximation, within the relativistic
quasiparticle random phase approximation \cite{PRN.03}. In the
Dirac-Hartree-BCS basis $\{|\psi_{k}^{(\eta)}\rangle\}$ it has the
following form:
\begin{equation}
\mathcal{R}_{\mu;k_{1}k_{2}}^{\eta}={\tilde{R}}_{k_{1}k_{2}}^{(0)\eta}%
(\Omega_{\mu})\sum\limits_{k_{3}k_{4}}\sum\limits_{\eta^{\prime}}\tilde
{V}_{k_{1}k_{4},k_{2}k_{3}}^{\eta\eta^{\prime}}\mathcal{R}_{\mu;k_{3}k_{4}%
}^{\eta^{\prime}}, \label{rqrpa}%
\end{equation}
where
$\tilde{V}_{k_{1}k_{4},k_{2}k_{3}}^{\eta\eta^{\prime}}=\tilde{V}_{k_{1}%
k_{4},k_{2}k_{3}}^{\eta,-\eta^{\prime},-\eta,\eta^{\prime}}$, since
we cut out the particle-hole components of the tensors in the
quasiparticle space.

In the diagrammatic expression of the amplitude (\ref{phires}) in
the upper line of the Fig.~\ref{fig1} the uncorrelated propagator
$\tilde{R}^{(0)\eta}_{k_1k_2}$ is represented by the two straight
nucleonic lines between the circles denoting emission and absorption
of a phonon by a single quasiparticle with the amplitude
$\gamma^{\eta_1\eta_2}_{\mu;k_1k_2}$. The approach to the amplitude
$\Phi(\omega)$ expressed by Eq. (\ref{phires}) represents a version
of first-order perturbation theory compared to RQRPA and the
amplitude $W(\omega)$ of Eq. (\ref{W-omega}) is the first-order
correction to the effective interaction ${\tilde V}$, because the
dimensionless matrix elements of the phonon vertices are such that
$\gamma^{\eta_1\eta_2}_{\mu;k_1k_2}/\Omega_{\mu}\ll 1$ in most
physical cases. The phonon-coupling term $\Phi$ generates
fragmentation of the excitation modes obtained in QRPA. In
particular, the high-frequency oscillations known as giant
resonances acquire their spreading width due to the term $\Phi$. In
the low-energy region below the neutron threshold of medium-mass
even-even nuclei this term is responsible solely for the appearing
strength. In the relativistic framework, the latter was confirmed
and extensively studied \cite{LRT.08,LRTL.09}, and verified by
comparison to experimental data
\cite{LRT.10,LRT.13,E.10,M.12,LVLS.14}. However, a comparison with
high-resolution experiments on the dipole strength below the neutron
threshold has revealed that, although the total strength and some
gross features of the strength are reproduced well, the fine
features are sensitive to truncation of the configuration space by
2q$\otimes$phonon configurations and further extensions of the
method are needed. Such an extension forms the content of the
subsequent sections.

\begin{figure}[ptb]
\begin{center}
\includegraphics[scale=1.05]{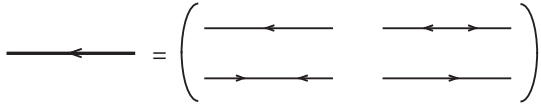}
\end{center}
\vspace{-0.5cm}
\caption{The 4-component Green's function in the diagrammatic
representation.}
\label{4-fg}%
\vspace{-0.3 cm}
\end{figure}

Before proceeding further, let us notice that the diagrammatic
equation of Fig. \ref{fig1} written, as in Ref. \cite{LRT.08}, for
the system with pairing correlations has the same form as that for
the normal (non-superfluid) system. The formal similarity of the
equations for normal and superfluid systems is achieved by the use
of the representation of the basis functions
$|\psi_{k}^{(\eta)}\rangle$ satisfying Eq.~(\ref{hb}). This basis is
a counterpart of the particle-hole basis of the conventional RPA in
which the (Q)RPA equations have the most simple and compact form. In
the representation of the functions $|\psi_{k}^{(\eta)}\rangle$ the
generalized superfluid mean-field Green function $\tilde{G}$ (often
called Gor'kov-Green's function) has a diagonal form and describes
the propagation of the quasiparticle with fixed energy. In this
diagonal representation the directions of the fermion lines of the
diagrams (of the type shown in Fig. \ref{fig1}) denote the positive-
or the negative-frequency components of the functions $\tilde{G}$.
The so-called backward-going diagrams, corresponding to the
ground-state correlations in the RQRPA, are not marked out in Fig.
\ref{fig1} though they are included in Eq. (\ref{respdir}). In the
coordinate representation, the non-diagonal Green function
$\tilde{G}$ for the quasiparticle has no definite energy. This Green
function can be represented by the 2$\times$2 block matrix shown in
Fig. \ref{4-fg}, see an extended discussion in Ref. \cite{LRT.13}.

In RQTBA based on the CDFT, the elimination of double counting
effects of the phonon coupling is performed by the subtraction of
the static contribution of the amplitude ${\Phi}$ from the residual
interaction in Eq. (\ref{W-omega}), since the parameters of the
underlying functional have been adjusted to experimental data for
ground states and include, thereby, the phonon coupling
contributions to the ground state. The subtraction of the phonon
coupling amplitude at zero frequency $\Phi(0)$ in Eq.
(\ref{W-omega}) acquires another important role for the excitations
which have an isoscalar dipole component, for example, the
electromagnetic dipole response. On the RQRPA level the elimination
of the $1^-$ spurious state is achieved by the use of a sufficiently
large 2q configuration space within a fully self-consistent approach
\cite{PRN.03}. In the extended theories based on the self-consistent
RQRPA the translational invariance can be restored by the
subtraction of the energy dependent interaction amplitude at zero
frequency. In the numerical implementation, due to numerical
inaccuracies, this state appears at a finite energy below 1 MeV
already in RQRPA, but due to the subtraction procedure, in extended
theories such as RQTBA and RQTBA-2 of Refs. \cite{LRT.10,LRT.13} the
accuracy of elimination of the spurious state is preserved. A
detailed description of the subtraction procedure which, in
addition, guarantees stability of solutions of the extended RPA
theories, is presented in Refs. \cite{Tse.13}. A similar procedure
is proposed in the higher-order RQTBA described in Section
\ref{multiconf}.

In practice, calculations for the response function (\ref{respdir})
are divided into two major steps. First, the BSE for the correlated
propagator $R^{(e)}(\omega)$
\bea R^{(e)\eta}_{k_1k_4,k_2k_3}(\omega) = {\tilde
R}^{(0)\eta}_{k_1k_2}(\omega)\delta_{k_1k_3}\delta_{k_2k_4} +
{\tilde R}^{(0)\eta}_{k_1k_2}(\omega)\times\nonumber\\
\times\sum\limits_{k_5k_6}\Bigl[{\Phi}^{\eta}_{k_1k_6,k_2k_5}(\omega)-
{\Phi}^{\eta}_{k_1k_6,k_2k_5}(0)\Bigr]
R^{(e)\eta}_{k_5k_4,k_6k_3}(\omega) \label{respdir2} \eea
is solved in the Dirac-Hartree-BCS basis. Second, the BSE for the
full response function $R(\omega)$ \bea
R_{k_{1}k_{4},k_{2}k_{3}}^{\eta\eta^{\prime}}(\omega) =
R^{(e)\eta}_{k_1k_4,k_2k_3}(\omega)\delta^{\eta\eta^{\prime}} +
\nonumber\\ + \sum\limits_{k_{5}k_{6}k_{7}k_{8}}
R^{(e)\eta}_{k_1k_6,k_2k_5}(\omega)
\sum\limits_{\eta^{\prime\prime}}{\tilde V}_{k_{5}k_{8}%
,k_{6}k_{7}}^{\eta\eta^{\prime\prime}}R_{k_{7}k_{4},k_{8}k_{3}}%
^{\eta^{\prime\prime}\eta^{\prime}}(\omega),\nonumber\\
\label{respdir1}%
\eea
where
\be R_{k_{1}k_{4},k_{2}k_{3}}^{\eta\eta^{\prime}}(\omega)=
R_{k_{1}k_{4},k_{2}k_{3}}^{\eta,-\eta^{\prime},-\eta,\eta^{\prime}}(\omega),
\ee
%
is solved either in the DHBCS or in the momentum-channel
representation which is especially convenient because of the
structure of the one-boson exchange interaction. The details are
given in Appendix C of Ref. \cite{LRT.08}.

\section{Extended RQTBA: the next order}

The first extension of the RQTBA described above used the idea
proposed in Ref.~\cite{Tse.07} and is based on the factorization of
Eq. (\ref{phires}): the uncorrelated propagator ${\tilde
R}^{(0)\eta}$ in Eq. (\ref{phires}) is replaced by the positive-
($\eta=+1$) or the negative- ($\eta=-1$) frequency part of a
correlated one. The first order approximation to a correlated
propagator is the RQRPA, in which a two-quasiparticle pair scatters
via a quasi-bound phonon configuration. As a result, two-phonon
configurations appear in the amplitude $\Phi(\omega)$, as it is
described in detail in Ref. \cite{LRT.13}. This two-phonon version
of the RQTBA, RQTBA-2, contains, by definition, more correlations
than the original RQTBA truncated by 2q$\otimes$phonon
configurations, but it is still on the same two-particle-two-hole
(2p2h) level of configuration complexity.

\begin{figure}
\begin{center}
\includegraphics*[scale=0.4]{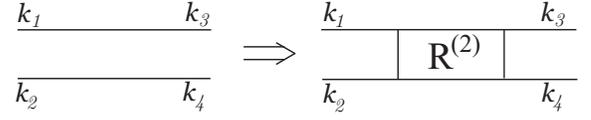}
\end{center}
\caption{Replacement of the uncorrelated two-nucleon propagator by
the correlated one. }
\label{subst}%
\end{figure}

\begin{figure*}[ptb]
\begin{center}
\includegraphics*[scale=0.8]{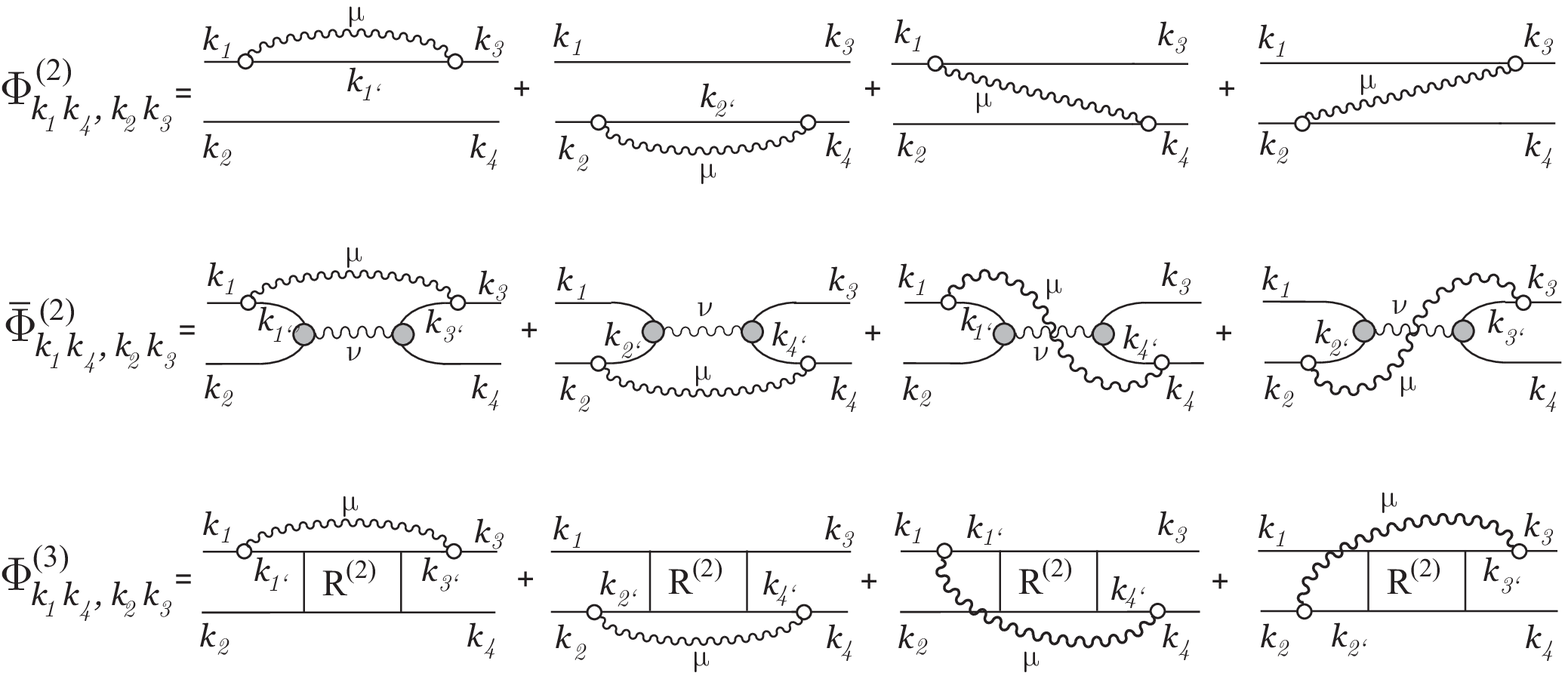}
\end{center}
\caption{Hierarchy of the quasiparticle-phonon coupling amplitudes:
2q$\otimes$phonon amplitude $\Phi = \Phi^{(2)}$ of the leading-order
QVC (compare to the bottom line of Fig. \ref{fig1}), the 2-phonon
amplitude ${\bar\Phi}^{(2)}$ and the 2q$\otimes$2phonon amplitude
$\Phi^{(3)}$ with the correlated intermediate two-quasiparticle
propagator $R^{(2)}$.}
\label{mph}%
\end{figure*}

Here we make another step forward with introducing correlations
inside the amplitude $\Phi(\omega)$. But now we go beyond the 2p2h
configurations. In order to keep the notations consistent with those
used before, let us define:
\bea
\Phi^{(1)\eta}_{k_1k_4,k_2k_3}(\omega) = 0 \nonumber\\
\Phi^{(2)\eta}_{k_1k_4,k_2k_3}(\omega) =
\Phi^{\eta}_{k_1k_4,k_2k_3}(\omega)\nonumber\\
R^{e(1)\eta}_{k_1k_4,k_2k_3}(\omega) = {\tilde
R}^{(0)\eta}_{k_1k_2}(\omega)\delta_{k_1k_3}\delta_{k_2k_4}\nonumber\\
R^{e(2)\eta}_{k_1k_4,k_2k_3}(\omega) =
R^{(e)\eta}_{k_1k_4,k_2k_3}(\omega)\nonumber\\
R^{(1)\eta\eta\prime}_{k_1k_4,k_2k_3}(\omega) = {\tilde
R}^{(0)\eta}_{k_1k_2}(\omega)\delta_{k_1k_3}\delta_{k_2k_4}\delta^{\eta\eta\prime}\nonumber\\
R^{(2)\eta\eta\prime}_{k_1k_4,k_2k_3}(\omega) =
R^{\eta\eta\prime}_{k_1k_4,k_2k_3}(\omega).
\eea
Now the response function $R^{(2)}$ of the conventional RQTBA
substitutes the uncorrelated intermediate propagator and, instead of
the amplitude $\Phi$ of Eq. (\ref{phires}), we have the new
amplitude $\Phi^{(3)}$:
\bea {\Phi}^{(3)\eta}_{k_1k_4,k_2k_3} (\omega) =
\nonumber\\
= \sum\limits_{k_5k_6,k_{5^{\prime}}k_{6^{\prime}}\mu}
\zeta^{\,\mu\eta}_{k_1k_2;k_5k_6}\,
R^{(2)\eta}_{k_5k_{6^{\prime}},k_6k_{5^{\prime}}} (\omega -
\eta\,\Omega_{\mu})\times\nonumber\\ \times\zeta^{\,\mu\eta
*}_{\,k_3k_4;k_{5^{\prime}}k_{6^{\prime}}},
\label{phires2} \eea
where the response function $R^{(2)\eta}_{k_1k_4,k_2k_3}$ is the
solution of the equation:
\bea
R_{k_{1}k_{4},k_{2}k_{3}}^{(2)\eta}(\omega) =
R^{(e)\eta}_{k_1k_4,k_2k_3}(\omega) + \nonumber\\
+ \sum\limits_{k_{5}k_{6}k_{7}k_{8}}
R^{(e)\eta}_{k_1k_6,k_2k_5}(\omega)
{\tilde V}_{k_{5}k_{8}%
,k_{6}k_{7}}^{\eta\eta}R_{k_{7}k_{4},k_{8}k_{3}}%
^{\eta}(\omega),\nonumber\\
\label{resp2_nogsc}%
\eea
which is an analog of Eq. (\ref{respdir1}), but does not contain
ground state correlations. This simplification is made to exclude
'zigzag' diagrams from the amplitude $\Phi^{(3)}$ \cite{Tse.pc2015}.
By this substitution, we introduce 2q$\otimes$phonon correlations
into the intermediate two-quasiparticle propagators, i.e., in
diagrammatic language, we perform the operation shown in Fig.
\ref{subst}.
The analytic expression for the new QVC amplitude in terms of the
phonon vertices (\ref{gammas}) reads:
\bea {\Phi}^{(3)\eta}_{k_1k_4,k_2k_3} (\omega) = \nonumber\\
=
\sum\limits_{k_{1'},k_{3'};\mu}\gamma^{\eta}_{\mu;k_1k_{1'}}R^{(2)\eta}_{k_{1'}k_2,k_{3'}k_4}(\omega-\eta\Omega_{\mu})
\gamma^{\eta\ast}_{\mu;k_3k_{3'}} + \nonumber\\
+
\sum\limits_{k_{2'},k_{4'};\mu}\gamma^{\eta}_{\mu;k_{2'}k_2}R^{(2)\eta}_{k_1k_{2'},k_3k_{4'}}(\omega-\eta\Omega_{\mu})
\gamma^{\eta\ast}_{\mu;k_{4'}k_4} - \nonumber \\
-
\sum\limits_{k_{1'},k_{4'};\mu}\gamma^{\eta}_{\mu;k_1k_{1'}}R^{(2)\eta}_{k_{1'}k_2,k_3k_{4'}}(\omega-\eta\Omega_{\mu})
\gamma^{\eta\ast}_{\mu;k_{4'}k_4} - \nonumber\\
-
\sum\limits_{k_{2'},k_{3'};\mu}\gamma^{\eta}_{\mu;k_{2'}k_2}R^{(2)\eta}_{k_1k_{2},k_{3'}k_4}(\omega-\eta\Omega_{\mu})
\gamma^{\eta\ast}_{\mu;k_3k_{3'}}, \label{phires2ex} \eea
where the four terms correspond to the four diagrams in the bottom
line of Fig. \ref{mph}. This Figure also illustrates the relation
between the QVC amplitudes in the conventional RQTBA, RQTBA-2 and in
the next-order approach.

\begin{figure}[ptb]
\vspace{-0.2cm}
\begin{center}
\includegraphics*[scale=0.49]{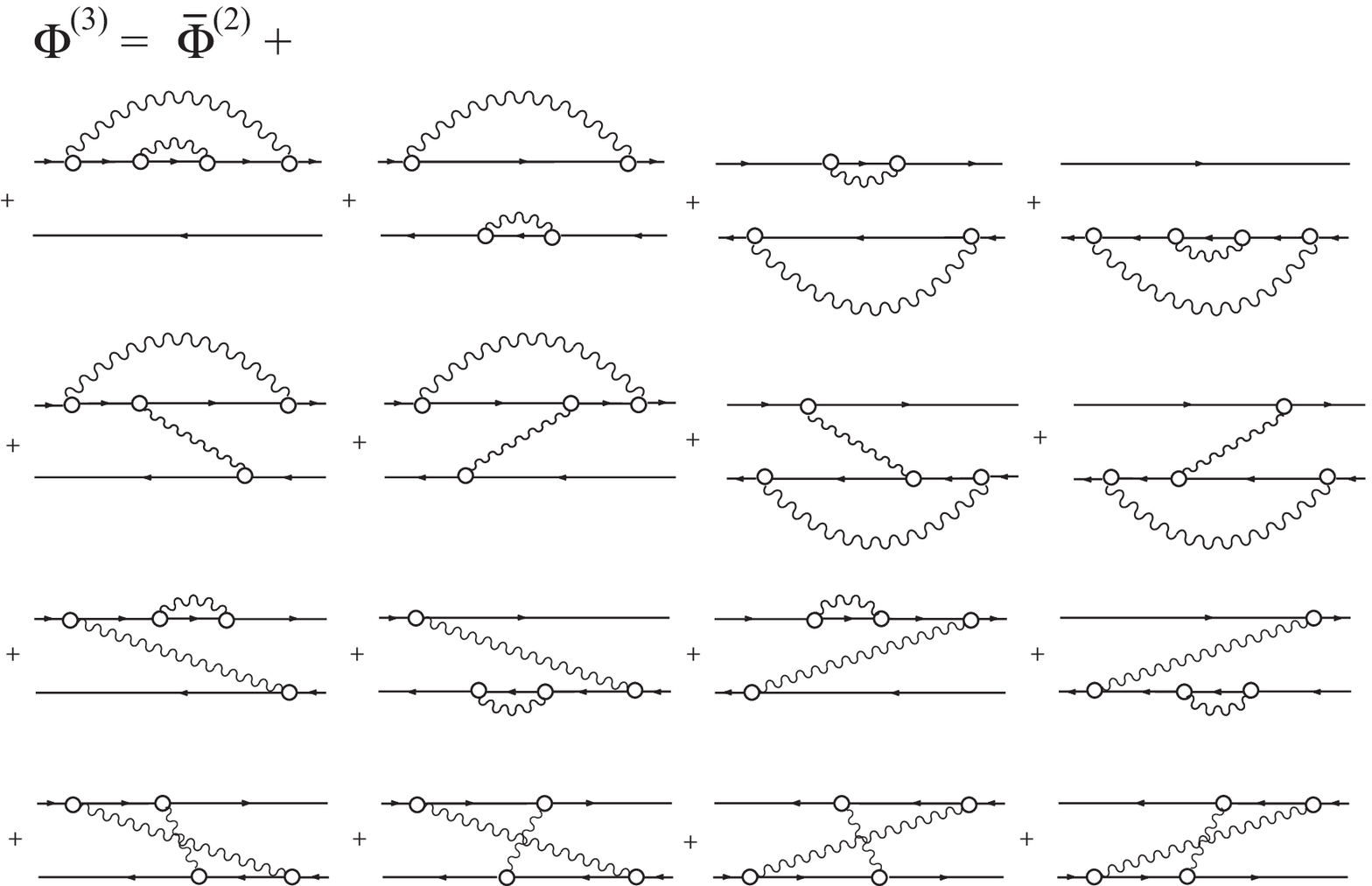}
\end{center}
\vspace{-0.5cm}
\caption{The time-ordered diagrams taken into account in the
extended RQTBA in the second order of the quasiparticle-vibration
coupling. }
\label{3ph}%
\end{figure}

In fact, the amplitude $\Phi^{(3)}$ contains the contributions of
the graphs shown in Fig. \ref{3ph}. However, the substitution shown
in Fig. \ref{subst} allows calculation of their contribution without
explicit calculations of the diagrams of Fig. \ref{3ph}. It is easy
to see that these terms contain 2q$\otimes$2phonon configurations
and thereby represent the next, three-particle-three-hole (3p3h),
level of configuration complexity, as compared to the RQTBA and
RQTBA-2. The two-phonon amplitude ${\bar\Phi}^{(2)}$ is also
contained in the amplitude $\Phi^{(3)}$, although approximately,
since the ground state correlations are neglected in Eq.
(\ref{resp2_nogsc}) \cite{Tse.pc2015}.

The amplitude $\Phi^{(3)}$ forms the kernel of the BSE for the
correlated propagator $R^{e(3)}$ taking into account 3p3h
correlations (to be compared to $R^{e(2)} = R^{(e)}$ which includes
2p2h ones):
\bea
R^{e(3)\eta}_{k_1k_4,k_2k_3}(\omega) = {\tilde
R}^{(0)\eta}_{k_1k_2}(\omega)\delta_{k_1k_3}\delta_{k_2k_4} +
{\tilde R}^{(0)\eta}_{k_1k_2}(\omega) \times\nonumber\\
\times\sum\limits_{k_5k_6}\Bigl[{\Phi}^{(3)\eta}_{k_1k_6,k_2k_5}(\omega)
- {\Phi}^{(3)\eta}_{k_1k_6,k_2k_5}(0)\Bigr]
R^{e(3)\eta}_{k_5k_4,k_6k_3}(\omega). \label{respe3}
\eea Analogously to the conventional RQTBA (\ref{respdir1}), the
equation for the full response function is formulated in terms of
the correlated propagator $R^{e(3)}$ as a free term and the static
effective interaction as a kernel: \bea
R_{k_{1}k_{4},k_{2}k_{3}}^{(3)\eta\eta^{\prime}}(\omega) =
R^{e(3)\eta}_{k_1k_4,k_2k_3}(\omega)\delta^{\eta\eta^{\prime}} +
\nonumber\\ + \sum\limits_{k_{5}k_{6}}
R^{e(3)\eta}_{k_1k_6,k_2k_5}(\omega)
\sum\limits_{k_{7}k_{8}\eta^{\prime\prime}}
{\tilde V}_{k_{5}k_{8}%
,k_{6}k_{7}}^{\eta\eta^{\prime\prime}}R_{k_{7}k_{4},k_{8}k_{3}}%
^{(3)\eta^{\prime\prime}\eta^{\prime}}(\omega),\nonumber\\
\label{respdir3}%
\eea
where the superscript '$(3)$' indicates that this response function
takes into account 3p3h configurations. Analogously to the
2q$\otimes$phonon RQTBA, the subtraction of the amplitude
$\Phi^{(3)}$ at zero frequency from the effective interaction in Eq.
(\ref{respdir3}) eliminates double counting of the static
contribution of phonon coupling effects.

\section{Higher-order correlations: multiphonon
coupling} \label{multiconf}

In principle, the procedure shown in Fig. \ref{subst} can be
repeated with the replacement $R^{(2)} \Rightarrow R^{(3)}$ to take
into account the next-order effects, and it can be continued until
convergence. Each iteration in this procedure will add another
correlated two-quasiparticle pair into the phonon coupling amplitude
$\Phi^{(n)}(\omega)$, resolving finer and finer features of the
response function. The latter means that in the model-independent
spectral expansion of the response function
\be
R^{(n)\eta\eta'}_{k_1k_4,k_2k_3} (\omega) =
\sum\limits_{\nu}\Bigl[\frac{\mathcal{R}^{(n)\eta\ast}_{\nu;k_1k_2}\mathcal{R}^{(n)\eta'}_{\nu;k_3k_4}}{\omega-\omega_{\nu}+i\delta}
-
\frac{\mathcal{R}^{(n)-\eta}_{\nu;k_2k_1}\mathcal{R}^{(n)-\eta'\ast}_{\nu;k_4k_3}}{\omega+\omega_{\nu}-i\delta}\Bigr]
\ee
more and more terms numbered by the index $\nu$ will appear with the
increase of $n$, so that the spectrum will become more and more
fragmented. In this way, the parameter $n$ establishes a hierarchy
of the excited states: larger $n$ numbers correspond to fine
structure while small $n$'s are responsible for gross structure of
the spectra. As it is shown below in Section \ref{coupled} by the
multipole expansion of $R^{(n)}$, each iteration introduces such a
geometrical factor into the kernel of the Bethe-Salpeter equation
for $R^{(n)}$, that contains some smallness providing a condition
for the convergence of the iterative procedure.

The chain of operator equations for the correlated propagator
$R^{e(n)}$, phonon coupling amplitude $\Phi^{(n)}$ and response
function $R^{(n)}$ looks as follows:
\be \begin{cases} R^{(1)}(\omega) = R^{e(1)}(\omega) = {\tilde R}^{0}(\omega)\\
R^{e(n)}(\omega) = {\tilde R}^{0}(\omega) + \\+{\tilde
R}^{0}(\omega)
\Bigl[\Phi[R^{(n-1)}(\omega)] - \Phi[R^{(n-1)}(0)]\Bigr]R^{e(n)}(\omega)\\
R^{(n)}(\omega) = R^{e(n)}(\omega) + R^{e(n)}(\omega){\tilde
V}R^{(n)}(\omega), \label{respn}
\end{cases} \ee
where $n > 1$ and the matrix elements of the amplitude
$\Phi[R^{(n-1)}(\omega)] = \Phi^{(n)}(\omega)$ are defined as:
\bea {\Phi}^{(n)\eta}_{k_1k_4,k_2k_3} (\omega) =
\nonumber\\
= \sum\limits_{k_5k_6,k_{5^{\prime}}k_{6^{\prime}}\mu}
\zeta^{\,\mu\eta}_{k_1k_2;k_5k_6}\,
R^{(n-1)\eta}_{k_5k_{6^{\prime}},k_6k_{5^{\prime}}} (\omega -
\eta\,\Omega_{\mu})\times\nonumber\\ \times\zeta^{\,\mu\eta
*}_{\,k_3k_4;k_{5^{\prime}}k_{6^{\prime}}}.
\label{phiresn} \eea
\begin{figure*}[ptb]
\begin{center}
\includegraphics*[scale=0.65]{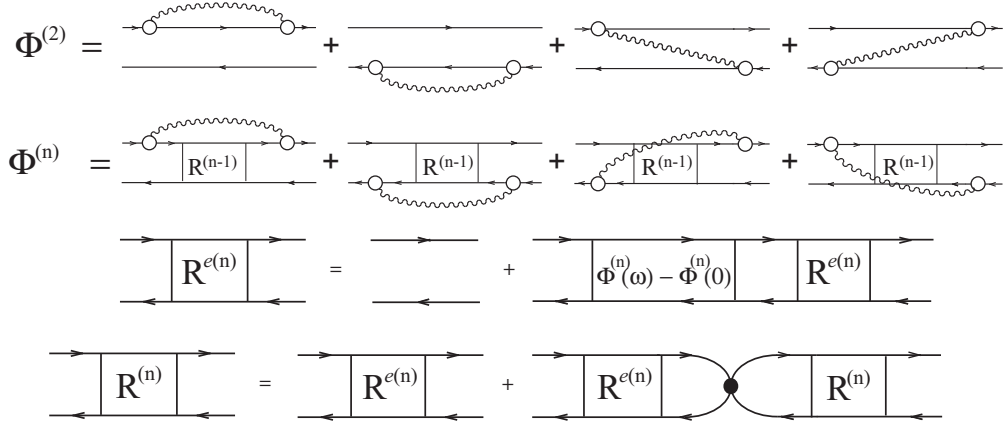}
\end{center}
\vspace{-0.5cm}
\caption{The diagrammatic representation of the iterative series for
the QVC amplitude and response function with multiphonon couplings:
The lowest order 2q$\otimes$phonon amplitude ${\Phi}^{(2)}$ of the
conventional phonon coupling model, the higher-order
2q$\otimes$(n-1)phonon amplitude ${\Phi}^{(n)}$ for $n > 2$;
Bethe-Salpeter equations in the ph-channel for the correlated
propagator $R^{e(n)}$ and for the response function $R^{(n)}$.
}
\label{fig2}%
\vspace{-0.3 cm}
\end{figure*}
In this context, the model, which was previously called RQTBA
('conventional' RQTBA), represents the second-order approach: if the
procedure (\ref{respn}) is truncated at $n = 2$, one would obtain
the conventional RQTBA of Ref. \cite{LRT.08}. The approach
(\ref{respn}) of the $n$-th order describes the nuclear response
function which includes couplings of two quasiparticles to up to
(n-1) phonons, or np-nh (2n quasiparticles, 2nq) configurations.
These effects are included in the time blocking approximation which
is, thereby, generalized to multiphonon coupling. Here only the
resonant part of the phonon coupling is taken into account, and the
so-called associated components introduced in Ref.
\cite{KTT.97,Tse.07} are neglected. Their quantitative role is known
to be minor for the spectral gross features, however, they represent
the ground state correlations caused by phonon dynamics and may
affect the fine structure of low-lying states \cite{KTT.97}.

In the diagrammatic language, the proposed solution (\ref{respn})
for the multiphonon response function is obtained by iteration of
the intermediate double-line (two-quasiparticle propagator) in the
QVC amplitude $\Phi$ of Fig. \ref{fig1}. This procedure is similar
to the method applied to the solution of the Dyson equation beyond
the leading order of QVC, when the single-quasiparticle Green
function entering the self-energy (\ref{sigmae}) is iterated.

The contribution of the terms with vertex corrections are known to
contain smallness, as compared to the line corrections, in analogy
to Migdal's theorem for electron-phonon systems \cite{Mig.58}. In
particular, in spherical nuclei all phonon-exchange terms which
represent vertex corrections (such as the last two terms in the
bottom line of Fig. \ref{fig1}), in their coupled form contain
6j-symbols which make these terms smaller than those associated with
self-energy insertions (first two terms in Fig. \ref{fig1}), see Eq.
(C4) of Ref. \cite{LRT.08}. Numerical calculations within the RQTBA
\cite{LRT.08,LRTL.09} have confirmed that the corrections to the
phonon vertices and frequencies beyond the RQRPA can be neglected
for the most important phonon modes. However, these corrections can
be, in principle, included into the iterative scheme (\ref{respn})
by extraction of the phonon vertices from the n-th order RQTBA
response function on each iteration.

In the next section the equations (\ref{respn}) are formulated in
the coupled form in the spherical DHBCS basis, which allows one to
adopt the approach for numerical calculations for finite nuclei.

\section{The multiphonon response function in the coupled form}
\label{coupled}

For practical calculations for finite nuclei, it is convenient to
formulate the equations (\ref{phires2ex}) - (\ref{phiresn}) in terms
of the reduced matrix elements with the transferred angular momentum
$J$, i.e. in the so-called coupled form. The reduced matrix elements
of the phonon-coupling amplitude $\Phi^{(n)}$ read:
\bea
{\Phi}^{(n)J,\eta}_{(k_1k_4,k_2k_3)}(\omega) =
\frac{(-1)^{j_1+j_2+j_3+j_4}}{2J+1}\sum\limits_{(\mu)J_e}\times\nonumber\\
\Bigl[%
\sum\limits_{(k_{1'}k_{3'})}\gamma^{\eta}_{(\mu;k_1k_{1'})}R^{(n-1)J_e,\eta}_{(k_{1'}k_2,k_{3'}k_4)}(\omega-\eta\Omega_{\mu})
\gamma^{\eta\ast}_{(\mu;k_3k_{3'})}\times\nonumber\\
\times \left\{
\begin{array}
[c]{ccc}%
J & J_{\mu} & J_e\\
j_{1'} & j_{2} & j_1%
\end{array}
\right\} \left\{
\begin{array}
[c]{ccc}%
J & J_{\mu} & J_e\\
j_{3'} & j_{4} & j_3%
\end{array}
\right\}+ \nonumber\\
+
\sum\limits_{(k_{2'}k_{4'})}\gamma^{\eta}_{(\mu;k_{2'}k_{2})}R^{(n-1)J_e,\eta}_{(k_{1}k_{2'},k_{3}k_{4'})}(\omega-\eta\Omega_{\mu})
\gamma^{\eta\ast}_{(\mu;k_{4'}k_{4})}\times\nonumber\\
\times \left\{
\begin{array}
[c]{ccc}%
J & J_{\mu} & J_e\\
j_{2'} & j_{1} & j_2%
\end{array}
\right\} \left\{
\begin{array}
[c]{ccc}%
J & J_{\mu} & J_e\\
j_{4'} & j_{3} & j_4%
\end{array}
\right\} - \nonumber\\
-
\sum\limits_{(k_{1'}k_{4'})}\gamma^{\eta}_{(\mu;k_1k_{1'})}R^{(n-1)J_e,\eta}_{(k_{1'}k_2,k_{3}k_{4'})}(\omega-\eta\Omega_{\mu})
\gamma^{\eta\ast}_{(\mu;k_{4'}k_{4})}\times\nonumber\\
\times \left\{
\begin{array}
[c]{ccc}%
J & J_{\mu} & J_e\\
j_{1'} & j_{2} & j_1%
\end{array}
\right\} \left\{
\begin{array}
[c]{ccc}%
J & J_{\mu} & J_e\\
j_{4'} & j_{3} & j_4%
\end{array}
\right\} - \nonumber\\
%
%
-
\sum\limits_{(k_{2'}k_{3'})}\gamma^{\eta}_{(\mu;k_{2'}k_{2})}R^{(n-1)J_e,\eta}_{(k_{1}k_{2'},k_{3'}k_{4})}(\omega-\eta\Omega_{\mu})
\gamma^{\eta\ast}_{(\mu;k_{3}k_{3'})}\times\nonumber\\
\times \left\{
\begin{array}
[c]{ccc}%
J & J_{\mu} & J_e\\
j_{2'} & j_{1} & j_2%
\end{array}
\right\} \left\{
\begin{array}
[c]{ccc}%
J & J_{\mu} & J_e\\
j_{3'} & j_{4} & j_3
\end{array}
\right\}\Bigr]. \nonumber\\
\label{phicoup} \eea
Here the indices in the brackets denote full sets of single-particle
quantum numbers with the excluded magnetic quantum numbers (total
angular momentum projections: $k_1 = \{(k_1),m_1 \}$. The correlated
propagator $R^{e(n)}$ is calculated in the symmetrized form:
\bea
R_{s(k_{1}k_{4},k_{2}k_{3})}^{e(n)J,\eta}(\omega)&=&{\tilde{R}}^{(0)J,\eta}%
_{s(k_{1}k_4,k_{2}k_3)}(\omega)+\nonumber\\
+{\tilde{R}}^{(0)\eta}_{(k_{1}k_{2})}(\omega)
\sum\limits_{(k_{6}\leq k_{5})}\Bigl[{\Phi}_{s(k_{1}k_{6},k_{2}%
k_{5})}^{(n)J,\eta}(\omega)&-&{\Phi}_{s(k_{1}k_{6},k_{2}k_{5})}^{(n)J,\eta}%
(0)\Bigr]\times\nonumber\\
&\times&R_{s(k_{5}k_{4},k_{6}k_{3})}^{e(n)J,\eta}(\omega),\nonumber\\
\label{correlated-propagator}%
\eea
where the matrix elements with the subscript "s" are symmetrized
with respect to one non-conjugated and one conjugated quasiparticle
index. Such a symmetrization allows a shortened summation in the
integral part of the Eq. (\ref{correlated-propagator}) and
simplifies, to some extent, the numerical calculations. The
symmetrized matrix elements of the mean field propagator ${\tilde
R}_s^{(0)}$ and of the phonon coupling amplitude $\Phi_s^{(n)}$ have
the following form:
\begin{eqnarray}
{\tilde{R}}^{(0)J,\eta}_{s(k_{1}k_{4},k_{2}k_{3})}(\omega)=%
{\tilde R}^{(0)\eta}_{(k_1k_2)}(\omega) \times\nonumber\\ \times
\bigl[\delta_{(k_1k_3)}\delta_{(k_2k_4)}+(-)^{\phi_{12}}%
\delta_{(k_1k_4)}\delta_{(k_2k_3)} \bigr],
\\
{\Phi}^{(n)J,\eta}_{s(k_1k_4,k_2k_3)}(\omega)=\frac{1}{1+%
\delta_{(k_3k_4)}} \times\nonumber\\ \times
\Bigl[{\Phi}^{(n)J,\eta}_{(k_1k_4,k_2k_3)}(\omega) +
(-)^{\phi_{12}}{\Phi}^{(n)J,\eta}_{(k_2k_4,k_1k_3)}(\omega)\Bigr],
\end{eqnarray}
with $\phi_{12}=J+l_1-l_2+j_1-j_2$. The BSE for the full response
function $R^{(n)}(\omega)$
\bea R_{(k_{1}k_{4},k_{2}k_{3})}^{(n)J,\eta\eta^{\prime}}(\omega)=
R_{s(k_{1}k_{4},k_{2}k_{3})}^{e(n)J,\eta}(\omega)\delta^{\eta\eta^{\prime}}
+\sum\limits_{(k_{6}\leq k_{5})}\times\nonumber\\
\times\sum\limits_{(k_{8}\leq
k_{7})\eta^{\prime\prime}}R_{s(k_{1}k_{6},k_{2}k_{5})}^{e(n)J,\eta}(\omega)
{\tilde V}^{J,\eta\eta^{\prime\prime}}_{(k_{5}k_{8},k_{7}k_{6})}R_{(k_{7}%
k_{4},k_{8}k_{3})}^{(n)J,\eta^{\prime\prime}\eta^{\prime}}(\omega)\nonumber\\
\label{responsej}%
\eea
is solved either in Dirac-Hartree-BCS or in momentum-channel
representations, see  Appendix C of Ref. \cite{LRT.08}.

Similar to the conventional RQTBA, the subtraction of the static
contribution of the phonon coupling amplitude $\Phi(\omega=0)$ from
the effective interaction should be performed to avoid double
counting effects of the quasiparticle-vibration coupling
\cite{Tse.13}. The subtraction can be done in the integral part
either of the equation for the correlated propagator $R_s^{e(n)}$ or
of the equation for the response function $R^{(n)}$. In the latter
case the subtraction acquires the meaning of renormalization of the
static effective interaction $\tilde V$. In this section, giving the
coupled-form expressions for these quantities, the subtraction is
performed in Eq. (\ref{correlated-propagator}) for the symmetrized
correlated propagator $R_s^{e(n)}$. This is more convenient
technically because the subtracted term has the same form as its
energy-dependent counterpart $\Phi(\omega)$.

After finding the response function $R^{(n)}$ of Eq.
(\ref{responsej}) it can be substituted to the Eq. (\ref{phicoup})
for the next-order QVC amplitude. In principle, the iterations can
be continued until the desired accuracy is reached. The closed
system of equations for the nuclear response function in the coupled
form presented in this section can be directly implemented for
numerical calculations. The case of $n=3$ is the first step beyond
the conventional RQTBA and includes 2q$\otimes$2phonon
configurations. It is clear from Eq. (\ref{phicoup}) that on the
large scale the 2q$\otimes$2phonon effects play a smaller role
compared to the 2q$\otimes$phonon ones of the RQTBA because of the
products of the two 6j-symbols in each term on the right hand side
of the Eq. (\ref{phicoup}), which are of geometrical nature.
Every next iteration contains an additional smallness of this
origin, thus providing conditions for convergence of the whole
procedure (\ref{respn}). The convergence will be examined in more
detail and verified by numerical implementation of the approach in
the future work.

The resulting linear response function $R^{(n)}(\omega)$ contains
all the information on the nuclear response to external one-body
operators. The observed spectrum of a nucleus excited by a
sufficiently weak external field $P$ as, for instance, an
electromagnetic field or a weak current, is described by the nuclear
polarizability which is a double convolution of the response
function with this field operator. The reduced matrix elements of
the external field operator have the following general coupled form:
\bea
P_{(k_{1}k_{2})}^{(p)J,\eta}=\sum\limits_{LS}
\frac{\delta_{\eta,1}+(-1)^{S}\delta_{\eta,-1}}{\sqrt{1 +
\delta_{(k_1k_2)}}}
\eta^{S}_{(k_1k_2)}\times\nonumber\\\times\langle(k_{1})\parallel
P^{(p)J}_{LS}\parallel(k_{2})\rangle,
\eea
where the index $(p)$ contains all possible quantum numbers, other
than those concretized here. The factors $\eta^{S}_{(k_1k_2)}$ are
determined by combinations of the quasiparticle occupation numbers
$u_k, v_k$ \cite{RS.80}:
\be
\eta_{(k_{1}k_{2})}^{S} =\frac{1}{\sqrt{1+\delta_{(k_{1}k_{2})}}%
}\Bigl(u_{k_{1}}v_{k_{2}}+(-1)^{S}v_{k_{1}}u_{k_{2}}\Bigr),
\label{eta}
\ee
obtained as a solution of Eq. (\ref{hb}). These combinations reflect
symmetrization in the integral part of Eq. (\ref{responsej}), which
enables one to take each $2q$-pair into account only once because of
the symmetry properties of the reduced matrix elements ${\tilde
V}^{J,\eta\eta^{\prime\prime}}_{(k_{5}k_{8},k_{7}k_{6})}$.

Nuclear polarizability in np-nh time blocking approximation reads:
\bea
\Pi^{(n)}_{P}(\omega) &=& \sum\limits_{(k_{2}\leq
k_{1})\eta}\sum\limits_{(k_{4}\leq
k_{3})\eta'}P_{(k_{1}k_{2})}^{(p)J,\eta\ast}\times
\nonumber\\
&\times&
R_{(k_{1}k_{4}%
,k_{2}k_{3})}^{(n)J,\eta\eta'}(\omega)P_{(k_{3}k_{4})}^{(p)J,\eta'},
\label{polariz}
\eea
and determines the microscopic strength function $S(E)$ as:
\be S(E) = -\frac{1}{\pi}\lim\limits_{\Delta\to +0}Im\
\Pi^{(n)}_P(E+i\Delta), \label{strf}\ee where a finite imaginary
part $\Delta$ of the energy variable is introduced in order to
obtain a smoothed envelope of the spectrum, if needed, which is
often the case for the correct comparison to experimental data with
limited resolution. Thus, Eqs. (\ref{polariz}), (\ref{strf}) relate
the obtained response function to experimental observations.

\section{Summary}

In this work the nuclear response theory is advanced beyond the
existing approaches in order to include the effects of multiphonon
coupling. The theory is formulated consistently for the covariant
framework based on meson-exchange nuclear forces, i.e. on the
effective quantum hadrodynamics as the underlying concept, although
it can be adopted for a non-relativistic framework based on one of
the modern density functionals.

While quantum hadrodynamics provides a fundamental description of
nuclear processes of short and medium range, there are long-range
correlations with ranges of the order of nuclear size, which can not
be described by an exchange of heavy and intermediate-mass mesons in
perturbative methods. In medium-mass and heavy nuclei, the
collective effects, such as low-lying vibrational modes, emerge as
'effective' degrees of freedom, which are in immediate relevance to
the energy scale of nuclear structure. An order parameter associated
with these degrees of freedom appears naturally in the covariant
response theory already on the RQRPA level, which helps to treat
them as effective quasi-bosonic fields responsible for the
long-range correlations. Their characteristics are computed
consistently from the meson-exchange interaction using, in the
leading-order approximation, RQRPA or equivalent techniques.
Thereby, the link between the short-range, medium-range, and
long-range correlations is established, that forms an essential part
of the covariant response theory.

The mathematical structure of the presented extension of nuclear
response theory is based on the idea of time-blocking which is
another key ingredient for this work. The time-blocking
approximation makes possible the selection of the most important
Feynman graphs containing quasiparticle-vibration coupling and their
subsequent non-perturbative treatment in a controlled way. The
developed method generalizes the time-blocking approximation to
multiphonon coupling and, thus, is capable of resolving fine details
of nuclear excitation spectra, which was quite limited in the
previous versions of the RQTBA. The generalized response theory
presented here is, thereby, a step forward to a more precise
solution of the nuclear many-body problem, which aims at a unified
description of both high-frequency collective states and low-energy
spectroscopy. The proposed approach is of a rather general character
and can be applied for other many-body Fermi systems with collective
degrees of freedom.

\section{Acknowledgement}
Discussions with V. Tselyaev are gratefully acknowledged. This work
was supported by US-NSF grants PHY-1204486 and PHY-1404343.


\end{document}